\documentclass[twocolumn,english,12pt]{iopart}
\usepackage[T1]{fontenc}
\usepackage[latin1]{inputenc}
\usepackage{graphicx}
\usepackage{amssymb}

\makeatletter
\usepackage{iopams}
\usepackage{setstack}

\usepackage{mathptmx}

\usepackage{cite}


\makeatother

\usepackage{babel}

\begin{document}

\title{Dynamics of entanglement between two free atoms with quantized motion}

\author{F Lastra and S Wallentowitz}

\address{Facultad de Física, Pontificia Universidad Católica de Chile, Casilla
306, Santiago 22, Chile}
\begin{abstract}
The electronic entanglement between two free atoms initially at rest
is obtained including the effects of photon recoil, for the case when
quantum dispersion can be neglected during the atomic excited-state
lifetime. Different from previous treatments using common or statistically
independent reservoirs, a continuous transition between these limits
is observed, that depends on the inter-atomic distance and degree
of localization. The occurrence of entanglement sudden death and birth
as predicted here deviates from the case where the inter-atomic distance
is treated classically by a static value. Moreover, the creation of
a dark state is predicted, which manifests itself by a stationary
entanglement that even may be created from an initially separable
state.
\end{abstract}


\maketitle

\section{Introduction\label{sec:Introduction}}

If a quantum system is brought into contact with another system with
much larger mode space, the system suffers decoherence \cite{joos},
which in Born-Markov approximation results as an exponential decay
of its coherences. Within this paradigm one may conclude that also
entanglement \cite{schroedi} should naturally suffer a degradation
with a similar behavior: an asymptotic decay to zero. However, recently
it has been shown that many exceptions from this expected behavior
exist, that show a vanishing of entanglement at finite time, including
its possible reappearance at later times. These features have been
denoted as entanglement sudden death (ESD) and birth (ESB), respectively
\cite{h3,diosi,dodd,dodd-2,atom-eberly,esd2-generic,eberly3-generic}. 

The occurrence of ESD and/or ESB has created increasing interest in
studying the dynamics of entanglement of various types of bipartite
systems, where the concurrence \cite{wooters} can be used as a measure
of entanglement. These systems include among others, systems of continuous
variables \cite{contvar,contvar2}, atoms in cavities \cite{atom-eberly,atom-yonac,atom-sainz,atom-cui},
cavity fields \cite{cavity,cavity-lopez}, spin chains \cite{spin-chain1,spin-chain},
and atoms in free space \cite{free-jamroz,tanas,free-orszag,nuestro}.
Also non-Markovian effects have been studied for the dynamics of bipartite
entanglement \cite{non-markovian,non-markovian2,leon}. Furthermore,
experiments have been performed to model \cite{almeida} or directly
measure \cite{almeida2} the dynamics of entanglement.

To best of our knowledge, at present, theoretical analysis has been
performed only within the framework of a master equation in the Born-Markov
approximation, where the atoms are supposed to have classical and
fixed positions \cite{lehmberg,lehmberg2}. In such a treatment \cite{tanas}
the dynamics of the concurrence has been shown to reveal in general
two different time scales, one given by the spontaneous decay rate
and the other by a distance-dependent collective decay rate. Furthermore,
the existence of a sequence of more than one ESD and one ESB has been
stated. In the limit of infinite inter-atomic distance, the well-known
results for two spin systems in statistically independent reservoirs
result \cite{atom-eberly}. However, in such a treatment the influence
of the relative quantized motion of the atoms is discarded. 

Physically, this may only be realized by situating the two atoms in
independent traps and in the limit where the trap frequency is $\omega\gg E_{{\rm r}}/\hbar$,
where $E_{{\rm r}}$ is the recoil energy of the atoms. In this Lamb--Dicke
limit the photon recoil is received by the trapping mechanism (Mössbauer
effect) and the atomic motion is unaffected. In all other cases, the
photon recoil affects the motion of the two atoms and in this way
the dynamics of the electronic entanglement between the atoms may
be modified. This is most relevant in the limit of free atoms. There
the relative quantized motion has to be included, in order to allow
the distinguishability between the atoms for properly addressing the
issue of their mutual entanglement \cite{nuestro}.\emph{ }We will
briefly review the arguments given in Ref. \cite{nuestro}:

Given two identical two-level atoms, initially at rest and distinguishable
by their differing positions, their distinguishability should be maintained
at least during the atom's excited-state lifetime $\tau_{0}=2\pi/\gamma_{0}$;
$\gamma_{0}$ being the natural linewidth of the electronic transition.
This condition requires that the quantum dispersion of the relative-position
wave-packet be sufficiently weak, so that during $\tau_{0}$ the rms
spread be \begin{equation}
\Delta r\ll\bar{r},\label{eq:distinguishability}\end{equation}
 where $\bar{r}$ is the mean inter-atomic distance. Due to quantum
dispersion, the initial rms spread $\Delta r_{{\rm i}}$ is enlarged
during the interval $\tau_{0}$ to\begin{equation}
\Delta r_{{\rm f}}=\Delta r_{{\rm i}}\sqrt{1+\left(\frac{l_{{\rm d}}}{\Delta r_{0}}\right)^{4}},\label{eq:dispersion}\end{equation}
where the dispersion length is $l_{{\rm d}}=\sqrt{h/\gamma_{0}m}$
with $h$ being Planck's constant and $m$ the atomic mass. 

Equation (\ref{eq:dispersion}) together with condition (\ref{eq:distinguishability})
leads to the condition\begin{equation}
\bar{r}\gg\Delta r_{{\rm i}}\gg\lambda_{0}\sqrt{\frac{E_{{\rm r}}}{\hbar\gamma_{0}}},\label{eq:dr-limits}\end{equation}
where $\lambda_{0}$ is the wavelength of the electronic transition
and the recoil energy is $E_{{\rm r}}=(\hbar k_{0})^{2}/2m$ with
$k_{0}=2\pi/\lambda_{0}$. As typically $E_{{\rm r}}\ll\hbar\gamma_{0}$,
the minimum distance and rms spread may still be much smaller than
$\lambda_{0}$. However, a zero rms spread is not permitted, as then
quantum dispersion would rapidly render the atoms indistinguishable.
Thus, to be consistent with the requirement of distinguishability
during $\tau_{0}$, a finite initial spread $\Delta r_{{\rm i}}$
within the limits (\ref{eq:dr-limits}) is required. 

This latter requirement implies the quantized treatment of the inter-atomic
motion, which includes the recoil effects of photon emissions. In
consequence, in this paper we present a fully quantized treatment
of both the electronic and relative motion. By use of a Wigner--Weisskopf
approach it is shown how the dynamics of the entanglement between
the atoms is affected by the photon recoil during spontaneous emissions.

The outline of the paper is as follows: In Sec. \ref{sec:Dynamics-of-two}
the interaction between the two atoms and the em field is treated
in the Wigner--Weisskopf approach and the dynamics of the relevant
probability amplitudes is solved. Using these solutions, in Sec. III
the electronic reduced density matrix is obtained and then used in
Sec. IV to derive the concurrence as a measure of entanglement. Finally,
in Sec. V a summary and conclusions are given.

\section{Dynamics of two two-level atoms interacting with the electromagnetic
field\label{sec:Dynamics-of-two}}

\subsection{Hamilton operator}

The system consists of two atoms of mass $m$ and with inter-atomic
distance vector $\hat{\vec{r}}$ and corresponding relative momentum
$\hat{\vec{p}}$. Each atom is composed of two electronic levels with
transition frequency $\omega_{0}$, described by the atomic pseudo
spin operator $\hat{\vec{S}}_{a}$ ($a=\pm1$). Both atoms interact
with the em field via their electronic transition dipole moment $\vec{d}$. 

The complete Hamilton operator for this system is given by \begin{equation}
\hat{H}=\int d^{3}k\sum_{\sigma}\hbar ck\hat{a}_{\vec{k},\sigma}^{\dagger}\hat{a}_{\vec{k},\sigma}+\sum_{a=\pm}\hbar\omega_{0}\hat{S}_{a,z}+\frac{\hat{p}^{2}}{m}+\hat{V}.\label{eq:H}\end{equation}
Here $\hat{a}_{\vec{k},\sigma}$ is the annihilation operator for
a photon in the plane-wave mode defined by the wave vector $\vec{k}$
and the polarization $\sigma$. The interaction between atoms and
em field reads\begin{equation}
\hat{V}=\int d^{3}k\sum_{\sigma}\sum_{a=\pm}\hbar\kappa_{\vec{k},\sigma}\hat{S}_{a,+}\hat{a}_{\vec{k},\sigma}e^{-ia\vec{k}\cdot\hat{\vec{r}}/2}+{\rm H.a.},\label{eq:V}\end{equation}
where the atom-photon coupling strength is given by the rate\begin{equation}
\kappa_{\vec{k},\sigma}=\vec{d}\cdot\vec{e}_{\vec{k},\sigma}E_{k}/\hbar,\label{eq:kappa}\end{equation}
with $\vec{e}_{\vec{k},\sigma}$ being the polarization unit vector
and with the vacuum electric-field strength \begin{equation}
E_{k}=\sqrt{\frac{\hbar ck}{16\pi^{3}\epsilon_{0}}}.\label{eq:E_k}\end{equation}

\subsection{Equations of motion for the probability amplitudes}

Assuming an initial vacuum state for the em field and an initial electronic
state of the atoms consisting of a superposition of only $|\uparrow\uparrow\rangle$
and $|\downarrow\downarrow\rangle$, the general form of the wave
function for the system is given by \begin{eqnarray}
|\Psi(t)\rangle & = & \int d^{3}r|\vec{r}\rangle_{{\rm rel}}\otimes\biggl\{\psi_{g}(\vec{r},t)|g\rangle+\psi_{e}(\vec{r},t)|e\rangle\nonumber \\
 & + & \int d^{3}k\sum_{\sigma}\sum_{a}\psi_{a;\vec{k},\sigma}(\vec{r},t)|a;\vec{k},\sigma\rangle\nonumber \\
 & + & \int d^{3}k\int d^{3}k^{\prime}\sum_{\sigma,\sigma^{\prime}}\psi_{\vec{k},\sigma;\vec{k}^{\prime},\sigma^{\prime}}(\vec{r},t)|\vec{k},\sigma;\vec{k}^{\prime},\sigma^{\prime}\rangle\biggr\}.\label{eq:|Psi>}\end{eqnarray}
Here the electronic-photonic states are defined as \begin{eqnarray}
|g\rangle & = & |\downarrow\downarrow\rangle_{{\rm el}}\otimes|{\rm vac}\rangle_{{\rm em}},\label{eq:|g>}\\
|e\rangle & = & |\uparrow\uparrow\rangle_{{\rm el}}\otimes|{\rm vac}\rangle_{{\rm em}},\label{eq:|e>}\\
|+;\vec{k},\sigma\rangle & = & |\downarrow\uparrow\rangle_{{\rm el}}\otimes\hat{a}_{\vec{k},\sigma}^{\dagger}|{\rm vac}\rangle_{{\rm em}},\label{eq:|k>}\\
|-;\vec{k},\sigma\rangle & = & |\uparrow\downarrow\rangle_{{\rm el}}\otimes\hat{a}_{\vec{k},\sigma}^{\dagger}|{\rm vac}\rangle_{{\rm em}},\\
|\vec{k},\sigma;\vec{k}^{\prime},\sigma^{\prime}\rangle & = & |\downarrow\downarrow\rangle_{{\rm el}}\otimes\hat{a}_{\vec{k},\sigma}^{\dagger}\hat{a}_{\vec{k}^{\prime},\sigma^{\prime}}^{\dagger}|{\rm vac}\rangle_{{\rm em}},\label{eq:|kk>}\end{eqnarray}
where the state (\ref{eq:|kk>}) is not normalized to unity in order
to contain the bosonic enhancement when two identical photons are
emitted. 

With the general form (\ref{eq:|Psi>}) the Laplace-transformed Schrödinger
equation, \begin{equation}
i\hbar s|\overline{\Psi}(s)\rangle=i\hbar|\Phi\rangle+\hat{H}|\overline{\Psi}(s)\rangle,\end{equation}
has to be solved with the initial condition $|\Psi(t=0)\rangle=|\Phi\rangle$.
The initial state $|\Phi\rangle$ is analogously expanded as Eq. (\ref{eq:|Psi>})
with the initial probability amplitudes being $\phi_{g}(\vec{r}),$
$\phi_{e}(\vec{r})$, whereas $\phi_{a;\vec{k},\sigma}(\vec{r})=\phi_{\vec{k},\sigma;\vec{k}^{\prime},\sigma^{\prime}}(\vec{r})=0$
because of the initial absence of photons. The corresponding Laplace-transformed
coupled equations for the probability amplitudes read then \begin{eqnarray}
\left(is+\omega_{0}\right)\overline{\psi}_{g}(\vec{r},s)=i\phi_{g}(\vec{r}),\label{eq:laplace1}\\
\left(is-\omega_{0}\right)\overline{\psi}_{e}(\vec{r},s)=i\phi_{e}(\vec{r})+\int d^{3}k\sum_{\sigma}\sum_{a}\kappa_{\vec{k},\sigma}\overline{\psi}_{a;\vec{k},\sigma}(\vec{r},s)e^{ia\vec{k}\cdot\vec{r}/2},\label{eq:laplace2}\\
\left(is-ck\right)\overline{\psi}_{a;\vec{k},\sigma}(\vec{r},s)=\kappa_{\vec{k},\sigma}^{\ast}\overline{\psi}_{e}(\vec{r},s)e^{-ia\vec{k}\cdot\vec{r}/2}+\int d^{3}k^{\prime}\sum_{\sigma^{\prime}}\kappa_{\vec{k}^{\prime},\sigma^{\prime}}\nonumber \\
\qquad\qquad\qquad\times\left[\overline{\psi}_{\vec{k},\sigma;\vec{k}^{\prime},\sigma^{\prime}}(\vec{r},s)+\overline{\psi}_{\vec{k}^{\prime},\sigma^{\prime};\vec{k},\sigma}(\vec{r},s)\right]e^{-ia\vec{k}^{\prime}\cdot\vec{r}/2},\label{eq:laplace3}\\
\left(is+\omega_{0}-ck-ck^{\prime}\right)\overline{\psi}_{\vec{k},\sigma;\vec{k}^{\prime},\sigma^{\prime}}(\vec{r},s)=\sum_{a}\left[\kappa_{\vec{k},\sigma}^{\ast}\overline{\psi}_{a;\vec{k}^{\prime},\sigma^{\prime}}(\vec{r},s)e^{ia\vec{k}\cdot\vec{r}/2}\right.\nonumber \\
\qquad\qquad\qquad+\left.\kappa_{\vec{k}^{\prime},\sigma^{\prime}}^{\ast}\overline{\psi}_{a;\vec{k},\sigma}(\vec{r},s)e^{ia\vec{k}^{\prime}\cdot\vec{r}/2}\right].\label{eq:laplace4}\end{eqnarray}
Here it was assumed that the resulting spectral width of the electronic
resonances is much larger than the Doppler shifts induced by the slow
relative motion. Under this assumption the kinetic energy of this
relative motion can be safely neglected.

Note, that with the appearance of the state $|\vec{r}\rangle_{{\rm rel}}$
in Eq. (\ref{eq:|Psi>}) a full quantized treatment of the relative
distance between the atoms is warranted and included by the use of
the correspondingly defined amplitudes. As a consequence, the neglection
of the kinetic energy in the equations of motion for the amplitudes,
cf. Eqs. (\ref{eq:laplace1})--(\ref{eq:laplace4}), is equivalent
to the Raman--Nath approximation in atomic-beam diffraction. There
the kinetic energy is neglected, but the photon recoil still generates
orders of diffraction of the passing atoms which correspond to dynamical
changes of their atomic momenta.

\subsection{Solution of the probability amplitudes}

From Eq. (\ref{eq:laplace1}) is can be readily seen that the ground-state
amplitude has the simple solution\begin{equation}
\psi_{g}(\vec{r},t)=\phi_{g}(\vec{r})e^{i\omega_{0}t}.\label{eq:sol-psi-g}\end{equation}
To obtain the solution for the excited-state amplitude $\overline{\psi}_{e}$,
we first insert Eq. (\ref{eq:laplace4}) into Eq. (\ref{eq:laplace3})
to obtain \begin{eqnarray}
\left(is-ck\right)\overline{\psi}_{a;\vec{k},\sigma}(\vec{r},s)=\kappa_{\vec{k},\sigma}^{\ast}\overline{\psi}_{e}(\vec{r},s)e^{-ia\vec{k}\cdot\vec{r}/2}\nonumber \\
\quad+\int d^{3}k^{\prime}\sum_{\sigma^{\prime}}\frac{2|\kappa_{\vec{k}^{\prime},\sigma^{\prime}}|^{2}}{is+\omega_{0}-c(k+k^{\prime})}\left[\overline{\psi}_{a;\vec{k},\sigma}(\vec{r},s)+\overline{\psi}_{-a;\vec{k},\sigma}(\vec{r},s)e^{-i\vec{k}^{\prime}\cdot\vec{r}}\right]\nonumber \\
\quad+\int d^{3}k^{\prime}\sum_{\sigma^{\prime}}\frac{2\kappa_{\vec{k}^{\prime},\sigma^{\prime}}\kappa_{\vec{k},\sigma}^{\ast}}{is+\omega_{0}-c(k+k^{\prime})}\nonumber \\
\qquad\times\left[\overline{\psi}_{a;\vec{k}^{\prime},\sigma^{\prime}}(\vec{r},s)e^{ia(\vec{k}-\vec{k}^{\prime})\cdot\vec{r}/2}+\overline{\psi}_{-a;\vec{k}^{\prime},\sigma^{\prime}}(\vec{r},s)e^{-ia(\vec{k}+\vec{k}^{\prime})\cdot\vec{r}/2}\right].\label{eq:psi-ak}\end{eqnarray}
 Assuming now the form\begin{equation}
\overline{\psi}_{a;\vec{k},\sigma}(\vec{r},s)=\overline{\psi}_{e}(\vec{r},s)\chi_{a;\vec{k},\sigma}(\vec{r},s),\label{eq:trick}\end{equation}
and assuming that we may expand the required solution $\chi_{a;\vec{k},\sigma}$
in powers of the atom-photon interaction rate $\kappa_{\vec{k},\sigma}$,\begin{equation}
\chi_{a;\vec{k},\sigma}(\vec{r},s)=\kappa_{\vec{k},\sigma}^{\ast}e^{-ia\vec{k}\cdot\vec{r}/2}\left[\chi_{a;\vec{k},\sigma}^{(1)}(\vec{r},s)+|\kappa_{\vec{k},\sigma}|^{2}\chi_{a;\vec{k},\sigma}^{(3)}(\vec{r},s)+\ldots\right],\label{eq:chi-series}\end{equation}
 we obtain from the first-order terms of Eq. (\ref{eq:psi-ak}) the
condition\begin{equation}
\chi_{a;\vec{k},\sigma}^{(1)}(\vec{r},s)=\frac{1}{is-ck},\label{eq:chi1}\end{equation}
and likewise from the third-order terms the condition\begin{eqnarray}
|\kappa_{\vec{k},\sigma}|^{2}\chi_{a;\vec{k},\sigma}^{(3)}(\vec{r},s) & = & 2\chi_{a;\vec{k},\sigma}^{(1)}(\vec{r},s)\int d^{3}k^{\prime}\sum_{\sigma^{\prime}}\frac{|\kappa_{\vec{k}^{\prime},\sigma^{\prime}}|^{2}}{is+\omega_{0}-c(k+k^{\prime})}\nonumber \\
 & \times & \left\{ \frac{1+e^{i(a\vec{k}-\vec{k}^{\prime})\cdot\vec{r}}}{is-ck}+\frac{1+e^{ia(\vec{k}-\vec{k}^{\prime})\cdot\vec{r}}}{is-ck^{\prime}}\right\} .\label{eq:chi3}\end{eqnarray}

Inserting then Eq. (\ref{eq:trick}) into Eq. (\ref{eq:laplace2})
and using the expansion (\ref{eq:chi-series}), the excited state
amplitude is obtained as \begin{equation}
\overline{\psi}_{e}(\vec{r},s)=f(\vec{r},s)\phi_{e}(\vec{r}),\label{eq:laplace-solution}\end{equation}
where the function is defined as\begin{equation}
f(\vec{r},s)=\frac{1}{s+i[\omega_{0}+\Omega(\vec{r},s)]},\label{eq:def-f}\end{equation}
with the complex-valued frequency shift being\begin{equation}
\Omega(\vec{r},s)=\Omega^{(1)}(\vec{r},s)+\Omega^{(3)}(\vec{r},s)+\ldots\end{equation}

The first-order term of this frequency shift results from $\chi^{(1)}$
and reads \begin{equation}
\Omega^{(1)}(\vec{r},s)=2\int d^{3}k\sum_{\sigma}\frac{|\kappa_{\vec{k},\sigma}|^{2}}{is-ck},\label{eq:def-Omega1}\end{equation}
whereas the third-order term results from $\chi^{(3)}$ as\begin{eqnarray}
\Omega^{(3)}(\vec{r},s) & = & 4\int d^{3}k\sum_{\sigma}\frac{|\kappa_{\vec{k},\sigma}|^{2}}{is-ck}\int d^{3}k^{\prime}\sum_{\sigma^{\prime}}\frac{|\kappa_{\vec{k}^{\prime},\sigma^{\prime}}|^{2}}{is+\omega_{0}-ck-ck^{\prime}}\nonumber \\
 & \times & \left(\frac{1}{is-ck}+\frac{1}{is-ck^{\prime}}\right)\left\{ 1+\cos[(\vec{k}-\vec{k}^{\prime})\cdot\vec{r}]\right\} .\label{eq:def-Omega3}\end{eqnarray}

We note that each occurrence of the factor $|\kappa_{\vec{k},\sigma}|^{2}$
in Eqs (\ref{eq:def-Omega1}) and (\ref{eq:def-Omega3}) will result
as the natural line width $\gamma_{0}$ of the atomic electronic transition,
which is defined as\begin{equation}
\gamma_{0}=\frac{d^{2}\omega_{0}^{3}}{3\pi\epsilon_{0}\hbar c^{3}}.\end{equation}
Therefore $\Omega^{(1)}(\vec{r},s)\propto\gamma_{0}$ and $\Omega^{(3)}(\vec{r},s)\propto\gamma_{0}^{2}$,
where $\gamma_{0}\ll\omega_{0}$, so that the pole in Eq. (\ref{eq:def-f})
is located very near to $s=-i\omega_{0}$ and $\Omega$ represents
a relative displacement of this pole of the order of $\epsilon=\gamma_{0}/\omega_{0}\ll1$.
Thus we may approximate this frequency shift by taking the limiting
value \begin{equation}
\Omega(\vec{r})\approx\lim_{is\to\omega_{0}}\Omega(\vec{r},s)=\lim_{is\to\omega_{0}}\Omega^{(1)}(\vec{r},s)+\mathcal{O}(\gamma_{0}^{2}).\label{eq:Omega-approx}\end{equation}
The limiting value is obtained by evaluating the residues of expression
(\ref{eq:def-Omega1}) as $\lim_{is\to\omega_{0}}\Omega^{(1)}(\vec{r},s)=-i\gamma_{0}$,
so that we may use the approximation\begin{equation}
\Omega\approx-i\gamma_{0}.\label{eq:Omega-approx-1}\end{equation}

In the time domain, the function $f(\vec{r},t)$ can thus be approximated
as \begin{equation}
f(\vec{r},t)\approx f(t)=\exp\left[-(i\omega_{0}+\gamma_{0})t\right],\label{eq:f1}\end{equation}
being independent of the inter-atomic distance vector $\vec{r}$.
Correspondingly the amplitude (\ref{eq:trick}) can be consistently
approximated as\begin{equation}
\overline{\psi}_{a;\vec{k},\sigma}(\vec{r},s)=\frac{\kappa_{\vec{k},\sigma}^{\ast}e^{-ia\vec{k}\cdot\vec{r}/2}}{is-ck}f(t)\phi_{e}(\vec{r}).\label{eq:solution-psi_ak}\end{equation}

Equations (\ref{eq:sol-psi-g}), (\ref{eq:laplace-solution}) and
(\ref{eq:solution-psi_ak}) together with the approximation (\ref{eq:f1})
constitute the required information to evaluate the reduced electronic
density operator of the system.

\section{Electronic entanglement between the two atoms}

\subsection{Reduced electronic density matrix}

The reduced density operator of the electronic systems of the two
atoms reads

\begin{equation}
\hat{\varrho}(t)=\int d^{3}r{\rm Tr}_{{\rm em}}\langle\vec{r}|\psi(t)\rangle\langle\psi(t)|\vec{r}\rangle.\end{equation}
In the standard basis $\{|\uparrow\uparrow\rangle,|\uparrow\downarrow\rangle,|\downarrow\uparrow\rangle,|\downarrow\downarrow\rangle\}$
it can written as the reduced electronic density matrix \begin{equation}
\begin{tabular}{l}
 \ensuremath{\varrho(t)=\left(\begin{array}{cccc}
\varrho_{e,e} & 0 & 0 & \varrho_{e,g}\\
0 & \sigma_{+,+} & \sigma_{+,-} & 0\\
0 & \sigma_{+,-}^{\ast} & \sigma_{-,-} & 0\\
\varrho_{e,g}^{\ast} & 0 & 0 & \varrho_{g,g}\end{array}\right),}\end{tabular}\label{eq:density-matrix}\end{equation}
 where the matrix elements are given by \begin{eqnarray}
\varrho_{e,e}(t) & = & \int d^{3}r|\psi_{e}(\vec{r},t)|^{2},\label{eq:rho_ee}\\
\varrho_{e,g}(t) & = & \int d^{3}r\psi_{e}(\vec{r},s)\psi_{g}^{\ast}(\vec{r},t),\label{eq:rho_eg}\\
\sigma_{a,b}(t) & = & \int d^{3}r\int d^{3}k\sum_{\sigma}\psi_{a;\vec{k},\sigma}(\vec{r},t)\psi_{b;\vec{k},\sigma}^{\ast}(\vec{r},t),\label{eq:sigma_ab}\end{eqnarray}
and due to the unit trace,\begin{equation}
\varrho_{g,g}(t)=1-\varrho_{e,e}(t)-\sum_{a=\pm}\sigma_{a,a}(t).\label{eq:rho_gg}\end{equation}

For the evaluation of these matrix elements we will assume initial
conditions where the relative motion of the atoms is factorized from
the electronic state, i.e.\begin{eqnarray}
|\phi_{e}(\vec{r})|^{2} & = & \varrho_{e,e}(0)w(\vec{r}),\\
|\phi_{g}(\vec{r})|^{2} & = & \varrho_{g,g}(0)w(\vec{r}),\\
\phi_{e}(\vec{r})\phi_{g}^{\ast}(\vec{r}) & = & \varrho_{e,g}(0)w(\vec{r}),\end{eqnarray}
where $w(\vec{r})$ is the probability density for the inter-atomic
distance vector $\vec{r}$.

Given the approximation (\ref{eq:f1}), the solutions for the excited
state population and of the outer off-diagonal matrix element become
then \begin{eqnarray}
\varrho_{e,e}(t) & = & \varrho_{e,e}(0)e^{-2\gamma_{0}t},\label{eq:rho-ee}\\
\varrho_{e,g}(t) & = & \varrho_{e,g}(0)e^{-2i\omega_{0}t-\gamma_{0}t}.\label{eq:rho-eg}\end{eqnarray}
Due to the two possible decay channels $|\uparrow\uparrow\rangle\Rightarrow|\uparrow\downarrow\rangle$
and $|\uparrow\uparrow\rangle\Rightarrow|\downarrow\uparrow\rangle$,
the decay of the excited state population occurs at twice the decay
rate of a single atom.

\subsection{Density matrix elements $\sigma_{a,b}$}

From Eq. (\ref{eq:sigma_ab}) the Laplace transformed diagonal elements
$\sigma_{\pm,\pm}(s)$ can be shown to be identical and of the form\begin{equation}
\sigma_{\pm,\pm}(s)=\rho_{e,e}(0)\overline{\int_{0}^{s}ds^{\prime}f(\vec{r},s-s^{\prime})f^{\ast}(\vec{r},s^{\prime})I\left(s-s^{\prime},s^{\prime}\right)},\label{eq:sigma-diag}\end{equation}
where the averaging is done over the probability density $w(\vec{r})$
and the integral $I$ is given by ($s\geq p$) \begin{equation}
I(s,p)=\int d^{3}k\sum_{\sigma}\frac{|\kappa_{\vec{k},\sigma}|^{2}}{(ck-is)(ck+ip)}.\label{eq:I}\end{equation}
After performing the sum over polarizations and integrating over spherical
angles of the wave vector, this integral becomes%
\footnote{Note that the divergence of this integral is removed by the fact that
it is integrated over $s$ and $p$, as given in Eq. (\ref{eq:sigma-diag}).%
}\begin{equation}
I(s,p)=\frac{\gamma_{0}}{2\pi\omega_{0}^{3}}\int_{0}^{\infty}d\omega\frac{\omega^{3}}{(\omega-is)(\omega+ip)}.\label{eq:I-2}\end{equation}

Using Eqs (\ref{eq:f1}) and (\ref{eq:I-2}), the matrix elements
(\ref{eq:sigma-diag}) can be rewritten as\begin{equation}
\sigma_{\pm,\pm}(s)=\frac{\gamma_{0}}{2\pi\omega_{0}^{3}}\rho_{e,e}(0)\int_{0}^{\infty}d\omega\omega^{3}\int_{0}^{s}ds^{\prime}g(\omega,s-s^{\prime})g^{\ast}(\omega,s^{\prime}),\label{eq:sigma-diag-1}\end{equation}
where the new function $g$ is given by \begin{equation}
g(\omega,s)=\frac{f(s)}{\omega-is}=\frac{i}{(s+i\omega)(s+i\omega_{0}+\gamma_{0})},\end{equation}
where the definition (\ref{eq:def-f}) together with Eq. (\ref{eq:Omega-approx-1})
have been used. Thus, in the time domain the diagonal matrix elements
are \begin{equation}
\sigma_{\pm,\pm}(t)=\frac{\gamma_{0}}{2\pi\omega_{0}^{3}}\rho_{e,e}(0)\int_{0}^{\infty}d\omega\omega^{3}|g(\omega,t)|^{2}.\label{eq:sigma-diag-1-2}\end{equation}

The Laplace transformed centered off-diagonal matrix elements can
by obtained analogously by inserting Eq. (\ref{eq:solution-psi_ak})
into Eq. (\ref{eq:sigma_ab}). Both elements result identical and
read \begin{equation}
\sigma_{\pm,\mp}(s)=\rho_{e,e}(0)\overline{\int_{0}^{s}ds^{\prime}f(\vec{r},s-s^{\prime})f^{\ast}(\vec{r},s^{\prime})I_{\xi}\left(s-s^{\prime},s^{\prime},\tau_{r}\right)},\label{eq:sigma-pm1}\end{equation}
where the retardation time of light propagation between the atoms
is $\tau_{r}=r/c$. The integral is defined as\begin{equation}
I_{\xi}(s,p,\tau)=\int d^{3}k\sum_{\sigma}\frac{|\kappa_{\vec{k},\sigma}|^{2}e^{-ick\tau\cos\theta}}{(ck-is)(ck+ip)},\qquad(s\geq p).\label{eq:integral-I2}\end{equation}
with $\xi=\sin^{2}\angle(\vec{r},\vec{d})$, assuming the induced
electrical dipole moments of the two atoms to be parallel. After summing
over polarizations and integrating over the spherical angles of the
wavevector, this integral becomes\begin{equation}
I_{\xi}(s,p,\tau)=\frac{\gamma_{0}}{2\pi\omega_{0}^{3}}\int_{0}^{\infty}d\omega\omega^{3}\frac{\mu_{\xi}(\omega\tau)}{(\omega-is)(\omega+ip)},\qquad(s\geq p),\label{eq:integral-I3}\end{equation}
where the dissipative part of the dipole-dipole interaction pattern
is given by \cite{lehmberg,lehmberg2}\begin{equation}
\mu_{\xi}(x)=\frac{3}{2}\left[(3\xi-2)\left(\frac{\cos x}{x^{2}}-\frac{\sin x}{x^{3}}\right)+\xi\frac{\sin x}{x}\right].\label{eq:mu-def}\end{equation}

Inserting Eq. (\ref{eq:integral-I3}) into (\ref{eq:sigma-pm1}) and
using the approximation (\ref{eq:f1}) the off-diagonal elements result
as\begin{eqnarray}
\sigma_{\pm,\mp}(s) & = & \frac{\gamma_{0}}{2\pi\omega_{0}^{3}}\rho_{e,e}(0)\int_{0}^{\infty}d\omega\omega^{3}\overline{\mu_{\xi}(\omega r/c)}\nonumber \\
 & \times & \int_{0}^{s}ds^{\prime}g(\omega,s-s^{\prime})g^{\ast}(\omega,s^{\prime}).\label{eq:sigma-pm2-1}\end{eqnarray}
Thus, in the time domain the matrix elements (\ref{eq:sigma-pm2-1})
become\begin{equation}
\sigma_{\pm,\mp}(t)=\frac{\gamma_{0}}{2\pi\omega_{0}^{3}}\rho_{e,e}(0)\overline{\int_{0}^{\infty}d\omega\omega^{3}|g(\omega,t)|^{2}\mu_{\xi}(\omega r/c)}.\label{eq:sigma-pm2-2}\end{equation}

In both expressions, Eqs (\ref{eq:sigma-diag-1-2}) and (\ref{eq:sigma-pm2-2}),
the spectral weighting factor $|g|^{2}$ results as a Lorentzian profile
modulated by a damped harmonic oscillation,\begin{equation}
|g(\omega,t)|^{2}=\frac{\pi}{\gamma_{0}}P(\omega)\left\{ 1+e^{-2\gamma_{0}t}-2\cos[(\omega-\omega_{0})t]e^{-\gamma_{0}t}\right\} ,\label{eq:gg}\end{equation}
where the Lorentzian spectrum is given by\begin{equation}
P(\omega)=\frac{1}{\pi}\frac{\gamma_{0}}{(\omega-\omega_{0})^{2}+\gamma_{0}^{2}}.\end{equation}

Due to the Lorentzian, the cosine in Eq. (\ref{eq:gg}) contains the
argument $(\omega-\omega_{0})t\lesssim\gamma_{0}t$. For $\gamma_{0}t\ll1$
we may therefore approximate Eq. (\ref{eq:gg}) by \begin{equation}
|g(\omega,t)|^{2}\approx\frac{\pi}{\gamma_{0}}P(\omega)\left(1-e^{-\gamma_{0}t}\right)^{2}.\label{eq:gg-approx}\end{equation}
This approximation is consistent both with the fact that $|g(\omega,t=0)|^{2}=0$
and with the limiting value for $\gamma_{0}t\gg1$. Therefore, it
may be used also for large times without introducing large error. 

Using the approximation (\ref{eq:gg-approx}), the off-diagonal elements
finally result as\begin{equation}
\sigma_{\pm,\mp}(t)=\frac{\rho_{e,e}(0)}{4}\overline{A_{\xi}(r)}\left(1-e^{-\gamma_{0}t}\right)^{2},\label{eq:sigma-pm2-3}\end{equation}
where the spectrally averaged dissipative dipole-dipole pattern reads\begin{equation}
A_{\xi}(r)=\frac{2}{\omega_{0}^{3}}\int_{0}^{\infty}d\omega\omega^{3}P(\omega)\mu_{\xi}(\omega r/c).\label{eq:A-def}\end{equation}

Performing analogous approximations for the diagonal matrix elements
(\ref{eq:sigma-diag-1-2}) results in\begin{equation}
\sigma_{\pm,\pm}(t)=\frac{\rho_{e,e}(0)}{4}A_{\xi}(0)\left(1-e^{-\gamma_{0}t}\right)^{2}.\label{eq:sigma-diag-1-3}\end{equation}
As $A_{\xi}(0)=[1+(\gamma_{0}/\omega_{0})^{2}]^{3/2}\approx1$, the
latter diagonal elements become\begin{equation}
\sigma_{\pm,\pm}(t)\approx\frac{\rho_{e,e}(0)}{4}\left(1-e^{-\gamma_{0}t}\right)^{2}.\label{eq:sigma-diag-1-4}\end{equation}

Note, that the inner matrix elements, Eqs. (\ref{eq:sigma-pm2-3})
and (\ref{eq:sigma-diag-1-4}), have only one time scale, $\gamma_{0}^{-1}$,
at which they increase in time. This is different from the results
obtained for two atoms with classical fixed positions \cite{tanas}.
There, two time scales have been derived, where the first coincides
with $\gamma_{0}^{-1}$ and the second is $[\gamma_{0}\mu_{\xi}(k_{0}r_{{\rm cl}})]^{-1}$,
where $r_{{\rm cl}}$ is the classical fixed inter-atomic distance.

\subsection{Averaged dissipative dipole-dipole pattern}

For the extreme far field, where the mean interatomic distance $\bar{r}\gg r_{{\rm c}}$
with the critical distance being defined by \begin{equation}
r_{{\rm c}}=\left(\frac{\omega_{0}}{\gamma_{0}}\right)\lambda_{0},\end{equation}
in the average (\ref{eq:A-def}) the spectral width of the Lorentzian
generates in the argument of the dipole-dipole pattern $\mu_{\xi}$
a corresponding width $\Delta x\gg2\pi$. Thus, the $2\pi$-periodic
oscillation of $\mu_{\xi}$ will be effectively averaged over to result
in $\overline{A_{\xi}(r)}\approx0$. Thus, in this case the off-diagonal
matrix elements are \begin{equation}
\sigma_{\pm,\mp}(t)=0,\qquad(\bar{r}\gg r_{{\rm c}}).\end{equation}

In the opposite case, when the mean inter-atomic distance satisfies
the condition $\bar{r}\lesssim r_{{\rm c}}$, which still permits
very large distances between the atoms, the Lorentzian generates a
sufficiently small width to provide a non-vanishing average $\langle A_{\xi}\rangle$.
Under these circumstances we may take the dipole-dipole pattern at
the maximum $\omega_{0}$ of the Lorentzian, to obtain \begin{equation}
\overline{A_{\xi}(r)}\approx\overline{\mu_{\xi}(k_{0}r)}.\label{eq:A-approx}\end{equation}

As Eq. (\ref{eq:A-approx}) decays naturally when extending to the
far and extreme far field, we may use it as a general approximation
valid in the near, inductive, and far field --- including the extreme
far field. Thus, the off-diagonal matrix elements can be further approximated
as\begin{equation}
\sigma_{\pm,\mp}(t)\approx\frac{\rho_{e,e}(0)}{4}\bar{\mu}_{\xi}\left(1-e^{-\gamma_{0}t}\right)^{2}.\label{eq:sigma-pm2-4}\end{equation}

\begin{figure}
\begin{centering}
\includegraphics[width=1\columnwidth]{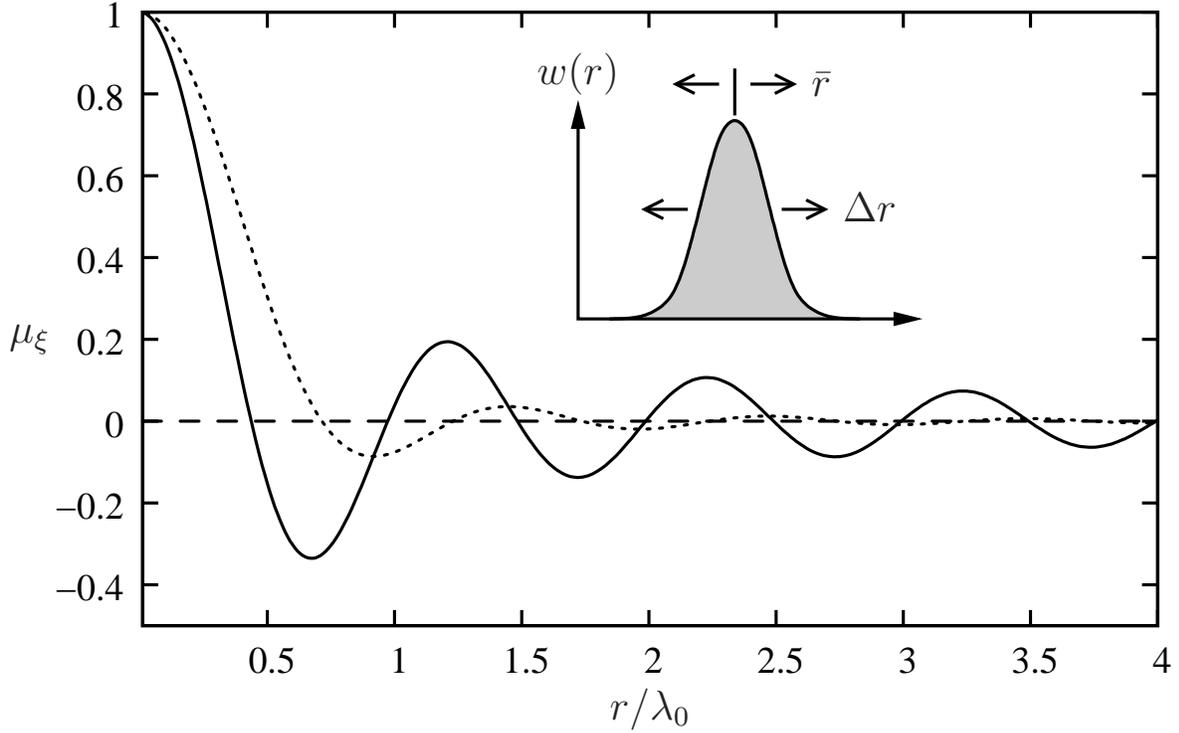}
\par\end{centering}

\caption{\label{fig:Dissipative-dipole-dipole-pattern}Dissipative dipole-dipole
pattern for atomic dipoles being perpendicular ($\xi=1$, solid curve)
and parallel ($\xi=0$, dotted curve) to the inter-atomic distance
vector. In the inset the dependence of the averaging over inter-atomic
distances on the mean distance $\bar{r}$ and rms spread $\Delta r$
is illustrated.}

\end{figure}

The average over the dissipative dipole-dipole pattern is illustrated
in Fig. \ref{fig:Dissipative-dipole-dipole-pattern}, where the function
$\mu_{\xi}(k_{0}r)$ is shown for the case when both atomic dipoles
are perpendicular (solid curve) or parallel (dotted curve) to the
distance vector between the atoms. As $-1<\mu_{\xi}\leq1$, also the
average is bound within the interval $-1<\bar{\mu}_{\xi}\leq1$. But
only for a strongly localized interatomic wavepacket that is positioned
near the anti-nodes of the oscillation of $\mu_{\xi}$ and in the
near field $\bar{r}\sim\lambda$, this average is non-vanishing. In
all other cases, e.g. far field, strongly localized near a node, or
weakly localized, this average vanishes. In this latter case the inner
off-diagonal elements of the reduced electronic density matrix are
$\sigma_{\pm,\mp}(t)=0$ for all times.

\section{Concurrence}

\subsection{Concurrence of the electronic subsystem of the two atoms}

For the density matrix (\ref{eq:density-matrix}) the concurrence
takes the form \begin{equation}
C(t)=2\max\{0,c_{1}(t),c_{2}(t)\},\end{equation}
where \begin{eqnarray}
c_{1}(t) & = & |\rho_{e,g}(t)|-\sqrt{\sigma_{+,+}(t)\sigma_{-,-}(t)},\label{eq:c1-def}\\
c_{2}(t) & = & |\sigma_{+,-}(t)|-\sqrt{\rho_{e,e}(t)\rho_{g,g}(t)}.\label{eq:c2-def}\end{eqnarray}
Using the results in Eqs (\ref{eq:rho-ee}), (\ref{eq:rho-eg}), (\ref{eq:sigma-diag-1-4}),
and (\ref{eq:sigma-pm2-4}), these expressions can be written as\begin{eqnarray}
c_{1}(\varepsilon) & = & |\rho_{e,g}(0)|\varepsilon-\frac{\rho_{e,e}(0)}{4}\left(1-\varepsilon\right)^{2},\label{eq:c1-2}\\
c_{2}(\varepsilon) & = & \frac{\rho_{e,e}(0)}{4}\bar{\mu}_{\xi}(k_{0}r)(1-\varepsilon)^{2}\nonumber \\
 & - & \varepsilon\sqrt{\rho_{e,e}(0)\left[1-\frac{\rho_{e,e}(0)}{2}+\rho_{e,e}(0)\varepsilon-\frac{3}{2}\rho_{e,e}(0)\varepsilon^{2}\right]},\label{eq:c2-2}\end{eqnarray}
where we defined $0<\varepsilon\leq1$ with \begin{equation}
\varepsilon=\exp(-\gamma_{0}t).\label{eq:x-def}\end{equation}

It can be shown that each of these functions, $c_{1}$ and $c_{2}$,
has exactly one zero within the interval $\varepsilon\in[0,1]$, $\varepsilon_{1}$
and $\varepsilon_{2}$, respectively. Moreover, these zeros are always
ordered temporally as \begin{equation}
\varepsilon_{1}\geq\varepsilon_{2},\end{equation}
which means that first the concurrence vanishes at time $t_{1}=-\ln\varepsilon_{1}/\gamma_{0}$
(ESD) and then revives at a later time $t_{2}=-\ln\varepsilon_{2}/\gamma_{0}$
(ESB). Thus exactly one single ESD and one single ESB exist for this
system. 

This behavior differs from that obtained within the framework of a
master equation where the atomic positions are considered as classical
and fixed values \cite{tanas}. In the latter case, more than a single
ESD and ESB may occur, depending on the initial conditions (see for
example Fig. 2 of Ref. \cite{tanas}). The difference is due to the
fact that here a different regime is considered. Namely, we allow
the atoms to move and to receive the photon recoil, whereas in Ref.
\cite{tanas} their positions are fixed (Lamb--Dicke limit).

Returning to the concurrence functions (\ref{eq:c1-2}) and (\ref{eq:c2-2}),
without loss of generality, the initial electronic state of the two
atoms may be written as the reduced density matrix\begin{equation}
\rho(0)=\left(\begin{array}{cccc}
p & 0 & 0 & q\sqrt{p(1-p)}\\
0 & 0 & 0 & 0\\
0 & 0 & 0 & 0\\
q\sqrt{p(1-p)} & 0 & 0 & 1-p\end{array}\right),\end{equation}
which for $q=1$ reduces to the pure state\begin{equation}
|\psi(0)\rangle=\sqrt{p}|\uparrow\uparrow\rangle+\sqrt{1-p}|\downarrow\downarrow\rangle.\end{equation}
With this choice it can be easily seen from Eqs (\ref{eq:c1-2}) and
(\ref{eq:c2-2}) that $c_{1}$ is a function of $p$ and $q$, whereas
$c_{2}$ depends only on $p$. Furthermore, only the function $c_{2}$
depends on the averaged dissipative dipole-dipole pattern $\bar{\mu}_{\xi}$.
As a consequence the two corresponding zeros are functions $\varepsilon_{1}=\varepsilon_{1}(p,q)$
and $\varepsilon_{2}=\varepsilon_{2}(p,\bar{\mu}_{\xi})$.

\begin{figure}
\begin{centering}
\includegraphics[width=1\columnwidth]{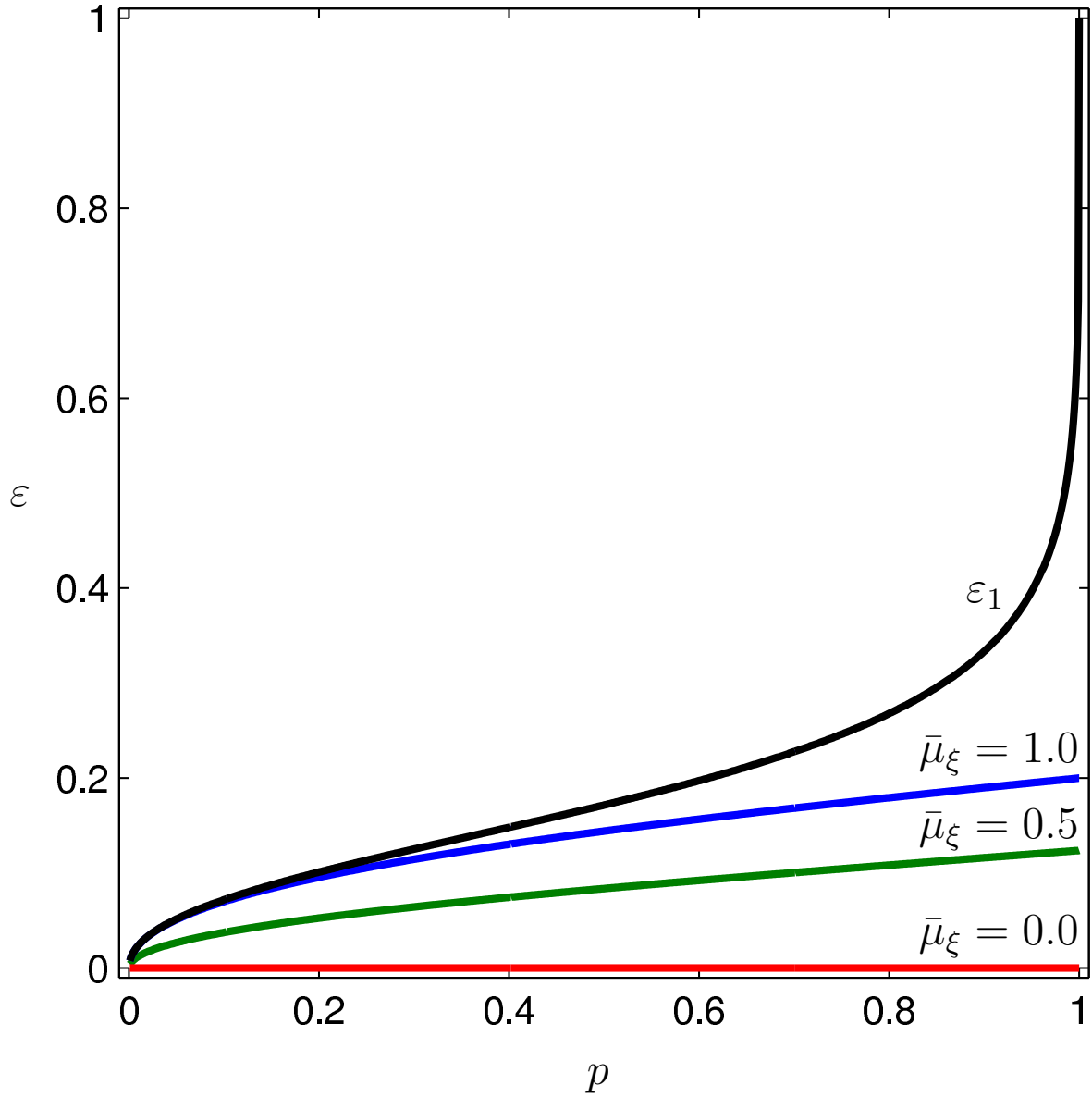}
\par\end{centering}

\caption{\label{fig:Zeros-of-the}Zeros of the concurrence functions $c_{1}$
(black) and $c_{2}$ (colored) as functions of the excited-state probability
$p$. The zeros of $c_{2}$ are plotted for values of $\bar{\mu}_{\xi}$
being: $1.0$ (blue), $0.5$ (green), $0.0$ (red).}

\end{figure}

For a pure state ($q=1)$ the dependence of the zeros on the excited-state
probability $p$ and on the mean dipole-dipole pattern $\bar{\mu}_{\xi}$
is shown in Fig. \ref{fig:Zeros-of-the}, where $\varepsilon_{1}$
is plotted in black and $\varepsilon_{2}$ in blue ($\bar{\mu}_{\xi}=1$),
green ($\bar{\mu}_{\xi}=1/2$), and red ($\bar{\mu}_{\xi}=0$). It
can be observed that for a given value of $p$, the time of disentanglement,
i.e. the time window between ESD and ESB, decreases with increasing
mean $\bar{\mu}_{\xi}$.

\begin{figure}
\begin{centering}
\includegraphics[width=1\columnwidth]{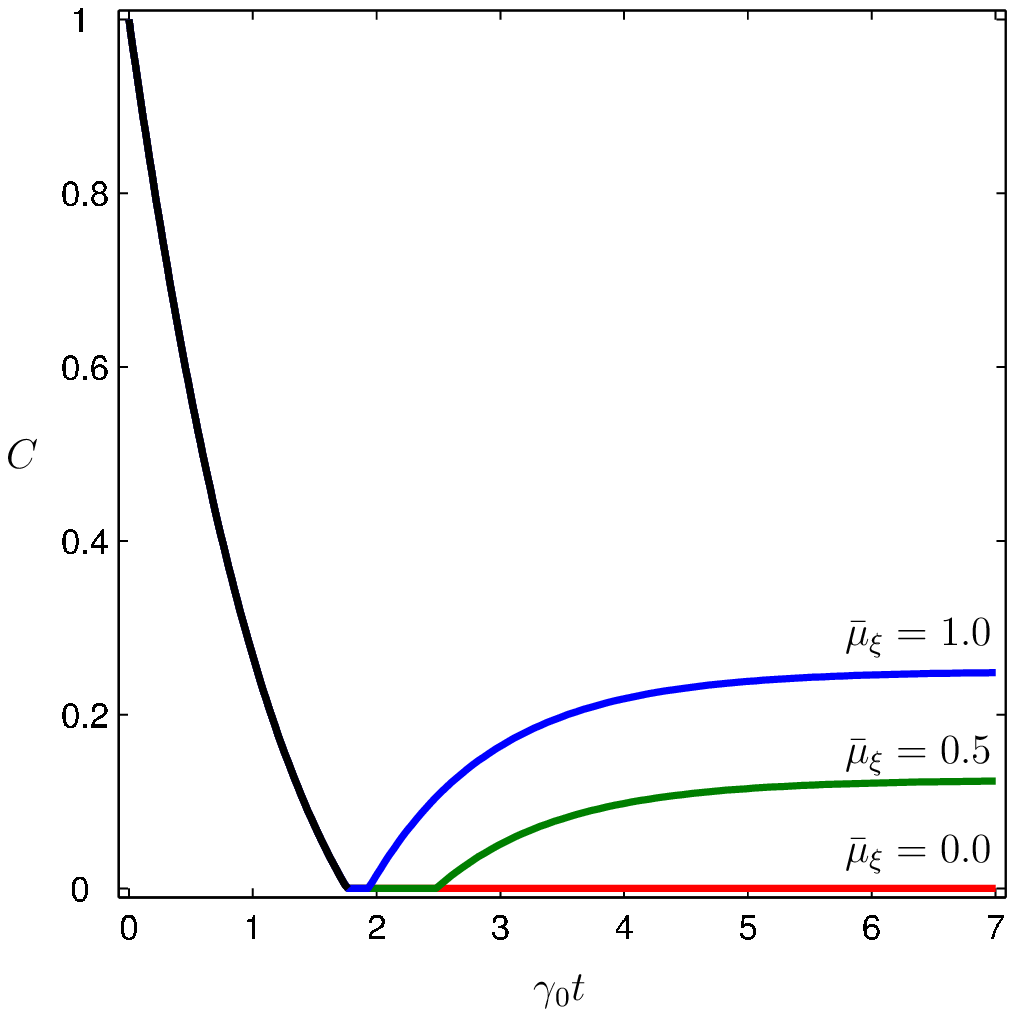}
\par\end{centering}

\caption{\label{fig:Concurrence-for-an}Concurrence for an initial pure maximally
entangled state ($q=1$, $p=1/2$) as a function of time for different
values of $\bar{\mu}_{\xi}$: 1.0 (blue), 0.5 (green), 0.0 (red).}

\end{figure}

A trivial case is obtained for $p=0$, where both atoms start in their
electronic ground states. In this case the concurrence vanishes at
all times. For $0<p<1$, which includes the case of initial maximal
entanglement ($p=1/2$), the concurrence first decays, ending with
an ESD, stays absent for a period of time, and then revives in the
form of an ESB, see Fig. \ref{fig:Concurrence-for-an}. Apart from
the case $\bar{\mu}_{\xi}=0$ (red curve) where the concurrence does
not revive --- which applies to far field, weak localization, and
strong localization near a node of $\mu_{\xi}$, the concurrence always
revives and reaches a stationary value. Thus, there is no asymptotic
decay to zero! 

\begin{figure}
\begin{centering}
\includegraphics[width=1\columnwidth]{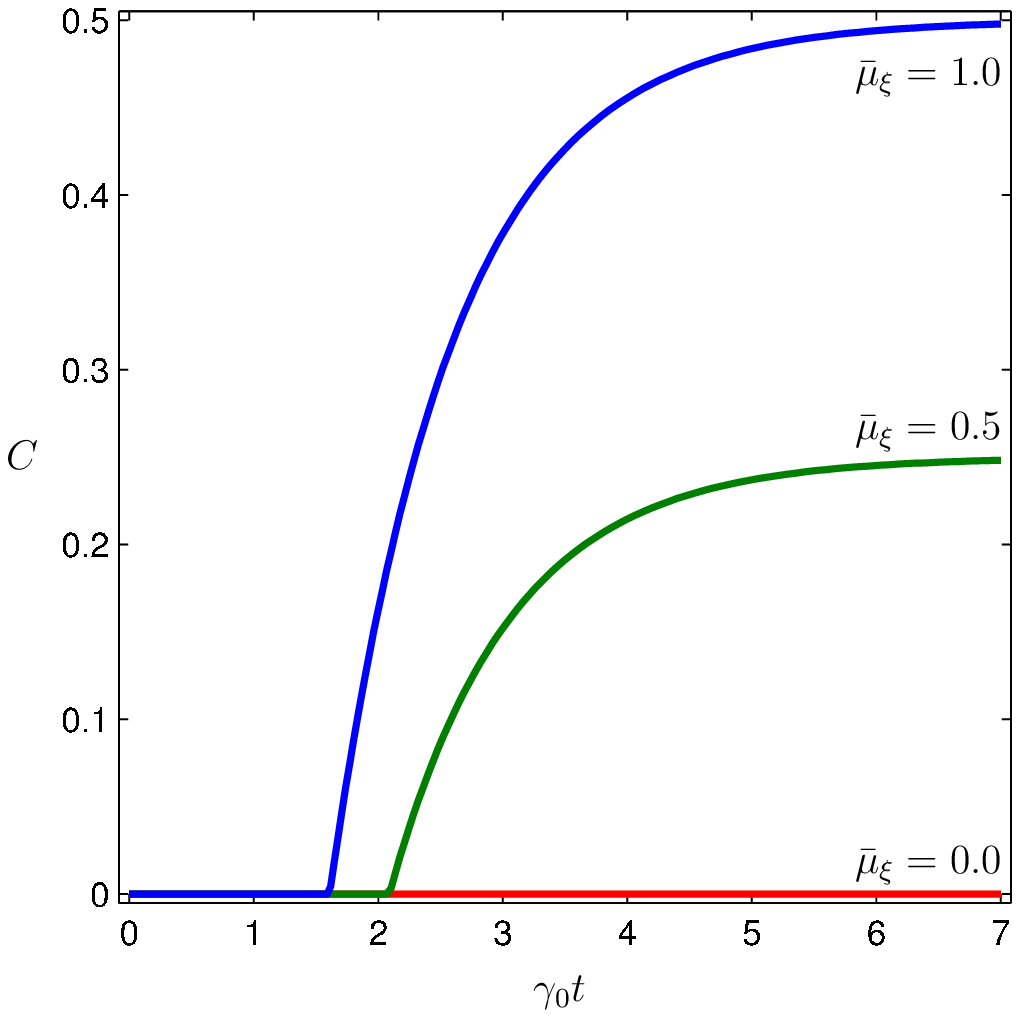}
\par\end{centering}

\caption{\label{fig:Concurrence-for-the}Concurrence for the initially separable
state $p=1$ ($q=1$), i.e. both atoms being initially excited, as
a function of time for different values of $\bar{\mu}_{\xi}$: 1.0
(blue), 0.5 (green), 0.0 (red).}

\end{figure}

The special case $p=1$, where initially both atoms are in their electronic
excited state and thus initially are not entangled, is of particular
interest. In this case a ESB is present without any previous ESD nor
entanglement, see Fig. \ref{fig:Concurrence-for-the}. This behavior
can be explained when considering that entanglement between the atoms
and the em field is created when one of the atoms emits a photon,
with the identity of the emitting atom being unknown. We will come
back to this issue in Sec. \ref{sub:Stationary-concurrence-due}.

\begin{figure}
\begin{centering}
\includegraphics[width=1\columnwidth]{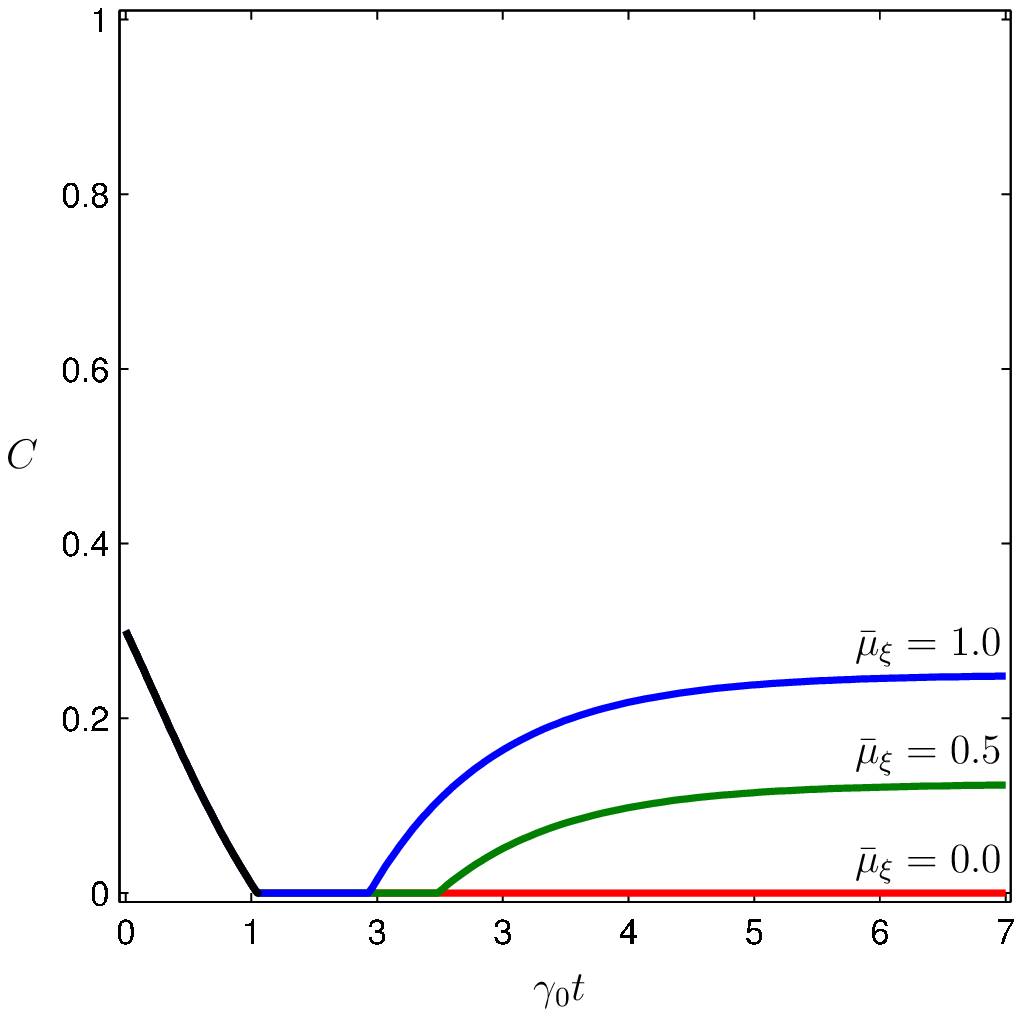}
\par\end{centering}

\caption{\label{fig:Concurrence-for-the-ini}Concurrence for the initially
mixed state $p=1/2$, $q=0.3$, i.e. a degraded maximally entangled
state, as a function of time for different values of $\bar{\mu}_{\xi}$:
1.0 (blue), 0.5 (green), 0.0 (red).}

\end{figure}

In the case of a mixed state, $c_{1}$ and thus also the zero $\varepsilon_{1}$,
are modified by the value of $q$. This corresponds to an obvious
modification of the initial value of the concurrence together with
a modification of the decay and thus ESD. However, the rebirth of
the concurrence is determined by the function $c_{2}$ and thus is
not affected, as can be seen when comparing Fig. \ref{fig:Concurrence-for-an}
with Fig. \ref{fig:Concurrence-for-the-ini}.

\subsection{Stationary concurrence due to a dark state\label{sub:Stationary-concurrence-due}}

For $t\to\infty$ the two functions, Eqs (\ref{eq:c1-2}) and (\ref{eq:c2-2})
reach the stationary values \begin{eqnarray}
\lim_{t\to\infty}c_{1}(t) & = & -\frac{\rho_{e,e}(0)}{4},\\
\lim_{t\to\infty}c_{2}(t) & = & \frac{\rho_{e,e}(0)}{4}\bar{\mu}_{\xi},\end{eqnarray}
so that there exists a stationary value for the concurrence,\begin{equation}
\lim_{t\to\infty}C(t)=\frac{\rho_{e,e}(0)}{2}\bar{\mu}_{\xi}.\label{eq:C-stat}\end{equation}

\begin{figure}
\begin{centering}
\includegraphics[width=1\columnwidth]{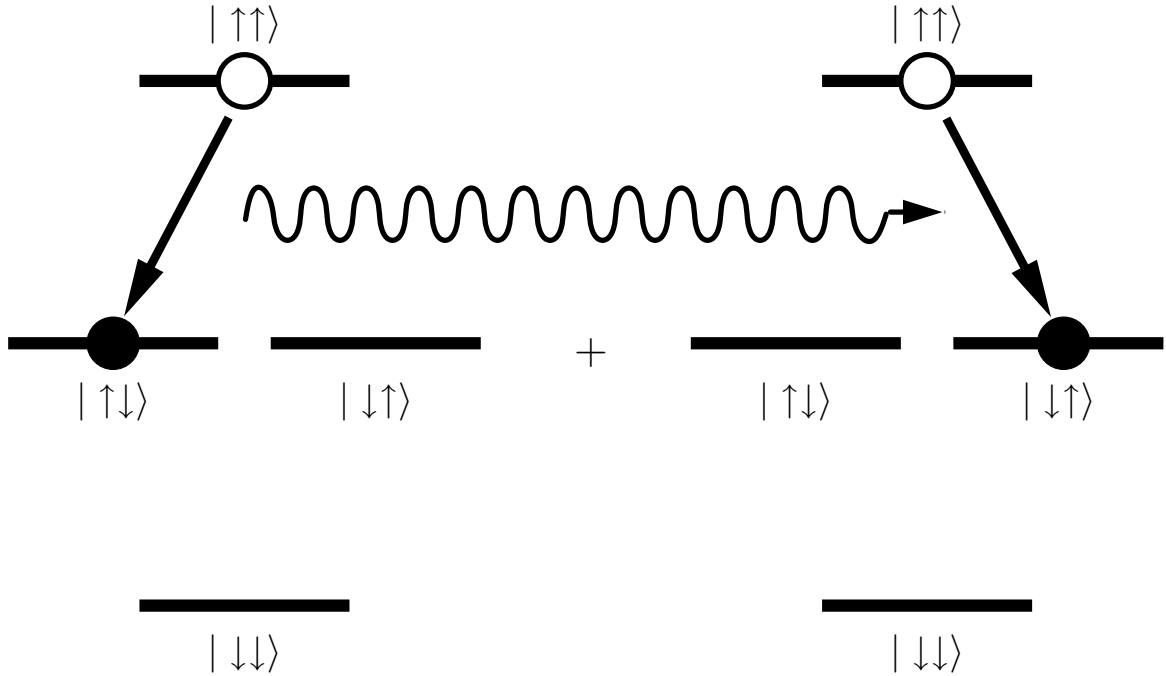}
\par\end{centering}

\caption{\label{fig:Dark-state-creation-from}Dark-state creation from state
$|\uparrow\uparrow\rangle$ by the emission of one photon: The system
evolves into a superposition of states $|\uparrow\downarrow\rangle|{\rm photon}\rangle$
and $|\downarrow\uparrow\rangle|{\rm photon}\rangle$.}

\end{figure}

This result might at first sight appear counter-intuitive, as it states
that there will be never any asymptotic decay to zero of the concurrence.
However, when considering the electronic-photonic and motional dynamics
it becomes clear that during the interaction of the atoms with the
em field a so-called {}``dark state'' is created. Starting from
the two atoms being excited, the system may evolve into a superposition
state where one photon has been emitted by either atom, see Fig. \ref{fig:Dark-state-creation-from}. 

This superposition state is an eigenstate of the Hamiltonian (\ref{eq:H})
when neglecting the quantum dispersion, i.e. discarding the kinetic
energy term. As a consequence, this state is stationary and thus a
further photon emission does not occur --- i.e. it is a dark state.
This stationary state manifests entanglement between all partitions.
If the emitted photon does not carry information on its emission center,
i.e. on the identity of the emitting atom, tracing over the em field
does not destroy the entanglement between the two atoms. Moreover,
if the associated photon recoil does not provide information on which
atom emitted the photon, tracing over the inter-atomic distance neither
diminishes the electronic entanglement between atoms. In such a case,
as the dark state is stationary, the concurrence reaches a stationary
value given by Eq. (\ref{eq:C-stat}). 

Contrary to this situation, when the emitted photon and/or the associated
recoil carry information on the identity of the emitting atom, after
tracing over the em field and inter-atomic distance, the reduced electronic
state does not reveal entanglement. This situation is realized in
various ways, all characterized by the condition $\bar{\mu}_{\xi}=0$.
For instance $\bar{\mu}_{\xi}=0$ if: (a) the atoms are weakly localized,
(b) strongly localized but near a node of $\mu_{\xi}$, or (c) in
the far field $\bar{r}\gg\lambda_{0}$. In the latter case one may
speak of two statistically independent reservoirs for the two atoms,
where our result is in agreement with Ref \cite{atom-eberly}.

Of course, as the kinetic-energy term in Eq. (\ref{eq:H}) has been
approximated, the discussed darkstate is not an exact eigenstate of
the full Hamiltonian (\ref{eq:H}). As a consequence, in a non-approximate
treatment the dispersion of the inter-atomic distance would lead to
a finite lifetime of this darkstate. However, this lifetime would
be much larger than the lifetime of the electronic excited state,
$2\pi/\gamma_{0}$ (cf. discussion in Sec. \ref{sec:Introduction}).
In this sense, our approximation is consistent for typical timescales
of electronic decay.

\section{Summary and Conclusions}

In summary it has been shown that different from the case of two atoms
with classical fixed positions, where more than one ESD and one ESB
may occur \cite{tanas}, for the case of two free atoms with quantized
relative motion and initially at rest, only one ESD and one ESB can
occur.  Furthermore, the inclusion of photon recoil permits the creation
of a dark state by spontaneous emission. This dark state  manifests
itself in a stationary electronic entanglement between the atoms,
that even may be created from an initially separable state. Whether
this stationary entanglement appears or not, depends crucially via
the photon recoil on the statistics of the inter-atomic distance.

We note, that a similar behavior has been observed in the non-Markovian
calculation of Ref. \cite{leon}. We believe that these phenomena
can be well explained within the framework of a dark state, as discussed
here.

\ack{}{FL and SW acknowledge support from FONDECYT 3085030, CONICYT-PBCT
(FL), and FONDECYT 1095214 (SW).}

\end{document}